\documentclass[sigconf]{acmart}

\usepackage{booktabs} 
\usepackage{multirow}
\usepackage{algorithm}
\usepackage[noend]{algpseudocode}

\usepackage{blindtext}
\usepackage{graphicx}
\usepackage{caption,subcaption}
\usepackage{wrapfig}
\usepackage{lipsum}

\usepackage{tabularx}
\usepackage{amssymb}

\usepackage{diagbox}
\usepackage{flushend}

\usepackage{amsmath,amsfonts,amsthm,bm}
\usepackage{stfloats}

\renewcommand\footnotetextcopyrightpermission[1]{} 
\begin{document}
\fancyhead{}
\title{SRCLock: SAT-Resistant Cyclic Logic Locking \\ for Protecting the Hardware}

\author{Shervin Roshanisefat}
\affiliation{%
  \institution{George Mason University}
  \streetaddress{}
  \city{}
  \state{}
  \postcode{}}
\email{sroshani@gmu.edu}  

\author{Hadi Mardani Kamali}
\affiliation{%
  \institution{George Mason University}
  \streetaddress{}
  \city{}
  \state{}
  \postcode{}}
\email{hmardani@gmu.edu}

\author{Avesta Sasan}
\affiliation{%
  \institution{George Mason University}
  \streetaddress{}  
  \city{}
  \state{}
  \postcode{}}
\email{asasan@gmu.edu}



\begin{abstract}
In this paper, we claim that cyclic obfuscation, when properly implemented, poses exponential complexity on SAT or CycSAT attack. The CycSAT, in order to generate the necessary cycle avoidance clauses, uses a pre-processing step. We show that this pre-processing step has to compose its cycle avoidance condition on all cycles in a netlist, otherwise, a missing cycle could trap the SAT solver in an infinite loop or force it to return an incorrect key. Then, we propose several techniques by which the number of cycles is exponentially increased with respect to the number of inserted feedbacks.  We further illustrate that when the number of feedbacks is increased, the pre-processing step of CycSAT faces an exponential increase in complexity and runtime, preventing the correct composition of loop avoidance clauses in a reasonable time before invoking the SAT solver. On the other hand, if the pre-processing is not completed properly, the SAT solver will get stuck or return incorrect key. Hence, when the cyclic obfuscation in accordance to the conditions proposed in this paper is implemented, it would impose an exponential complexity with respect to the number of inserted feedback, even when the CycSAT solution is used.
\end{abstract}



\maketitle
\section{Introduction}
The cost of building a new semiconductor fab was estimated to be \$5.0 billion in 2015, with large recurring maintenance costs \cite{DIGITIMES}\cite{6926108}, and sharply increases as technology migrates to smaller nodes. Due to the high cost of building, operating, managing, and maintaining state-of-the-art silicon manufacturing facilities, many major U.S. high-tech companies have been always fabless or went fabless in recent years (e.g., AMD, Broadcom, Marvell, Nvidia, Qualcomm, and Xilinx, to name a few). Thus, to reduce the fabrication cost, and for economic feasibility, most of the manufacturing and fabrication is pushed offshore \cite{DIGITIMES}. However, many offshore fabrication facilities are considered to be untrusted, which has raised concern over potential attacks in the manufacturing supply chain, with an intimate knowledge of the fabrication process, the ability to modify and expand the design prior to production, and an unavoidable access to the fabricated chips during testing. Hence, fabrication in untrusted fabs has introduced multiple forms of security threats from supply chain including that of overproduction, Trojan insertion, Reverse Engineering, Intellectual Property (IP) theft, and counterfeiting \cite{6926108}.

To prevent the adversaries from such attacks, researchers have proposed various obfuscation methods for hiding and/or locking the functionality of an netlist. However, the validity and strength of logic obfuscation to defend an IP against adversaries in the manufacturing supply chain was seriously challenged as researchers demonstrated that the de-obfuscation attacks leveraging satisfiability (SAT) solvers \cite{7140252}\cite{el2015integrated} combined with Signal Probability Skew (SPS) attacks \cite{7858346} could break the existing obfuscation schemes (both locking and camouflaging) in a relatively short time. Cyclic obfuscation \cite{Shamsi:2017:COC:3060403.3060458} was another approach that was considered as a defense mechanism against SAT solvers. However, this technique was later broken by CycSAT attack \cite{8203759}. The CycSAT added a pre-analysis step to the original SAT attack for detection and avoidance of cycles in the netlist during SAT attack. In this paper, we further investigate the CycSAT attack and illustrate its pre-processing step has to compose the cycle avoidance condition by traversing all cycles in a netlist. We illustrate that by having a small number of methodically constructed feedbacks in a netlist, an exponentially large number of simple and nested cycles could be generated in a netlist, and we propose two different techniques for building such behavior. Since a successful SAT attack on a cyclic circuit requires the avoidance clauses, and time it takes to generate such avoidance clauses has an exponential relation with the number of inserted feedbacks, the CycSAT attack faces exponential runtime at its processing step. Hence, when deploying the CycSAT, the complexity of the problem is not in the SAT solver step of the problem, but in its pre-processing step.

The rest of this paper is organized as follows. In section \ref{background} we cover the background on logic obfuscation. Then in section \ref{breakingCycSAt} we elaborate on the limitation of CycSAT and our approach for breaking the CycSAT. In section \ref{cyclock} we introduce our techniques for building an exponential relation between the number of feedbacks and the number of created cycles in a circuit. We also introduce three mechanisms for building a cyclic Boolean function to further increase the complexity of CycSAT pre-processing step. Our experimental results are summarized in section \ref{results}. Section \ref{conclusion} concludes the paper. 

\section{Background on Logic Obfuscation and SAT Attacks} \label{background}

Logic obfuscation is the process of hiding the functionality of an IP by building ambiguity or by implementing post manufacturing means of control and programmability into a netlist. Gate camouflaging and circuit locking are two of the widely explored obfuscation mechanisms \cite{6881480}\cite{7128395}\cite{7479225} for this purpose. A camouflaged gate is a gate that after reverse engineering (by means of delayering and lithography) could be mapped to any member of a possible set of gates or may look like one logic gate (e.g., AND), however functionally perform as another (e.g., XOR). In locking solutions, the functionality of a circuit is locked using a number of key inputs such that only when a correct key is applied, the circuit resumes its expected functionality. Otherwise, the correct function is hidden among many of the $2^K$ ($K$ being the number of keys) circuit possibilities. The claim raised by such obfuscation scheme was that to break the obfuscation, an adversary needs to try a large number of inputs and key combinations to extract the correct key, and the difficulty of this process increases exponentially as the number of keys and primary inputs increases. Hence, if enough gates are obfuscated, an adversary faces an unacceptably long time (claimed as years to decades) to break the obfuscation scheme. Note that the availability of scan chains (for DFT), allows an adversary to access combinational logic in each stage of a sequential circuit, load the desired input, execute the stage for one cycle, and readout the output.

The validity and strength of logic obfuscation to defend the IP against adversaries in the manufacturing supply chain was seriously challenged as researchers demonstrated that the satisfiability (SAT) solvers, when formulated according to Algorithm \ref{SAT_algoritm}, could break the obfuscation (both locking and camouflaging) in a matter of minutes as opposed to the promised claim of years and decades \cite{7140252}\cite{el2015integrated}. As illustrated in algorithm \ref{SAT_algoritm}, to employ a SAT attack, the obfuscated circuit is transformed into a circuit SAT problem, in which the SAT solver looks for an input value X for which the obfuscated circuit produces two different outputs for two different input keys. Such key is referred to as a \emph{Discriminating Input} $X_{DI}$. Each time a new $X_{DI}$ is found, the circuit SAT is updated to make sure that the next two keys that will be found in the next iteration of SAT solver invocation, produce the same output for all previously discovered $X_{DI}$. This is done by building a Discriminating Input Validation Circuit (DIVC) as illustrated in algorithm \ref{SAT_algoritm}. When the SAT solver can no longer find a $X_{DI}$, the DIVC circuit contains a complete set of discriminating inputs. At this point, any key that satisfies the DIVC (by calling a SAT solver on this circuit) is the key to the obfuscated circuit \cite{7140252}\cite{el2015integrated}. 

\begin{algorithm}
\caption{SAT Attack on Obfuscated Circuits \label{SAT_algoritm}}
\begin{algorithmic}[1]
\scriptsize
\State $DIVC = 1$;
\State $SAT_{circut} = C(X,K_1,Y_1) \wedge C(X,K_2,Y_2) \wedge (Y_1 \ne Y_2)$;
\While {$((X_{DI},K_1,K_2)\leftarrow SAT_F(SAT_{circut})=T)$} 
    \State $Y_f \leftarrow C_{BlackBox}(X_{DI})$; 
    \State $DIVC = DIVC \wedge C(X_{DI},K_1,Y_f) \wedge C(X_{DI},K_2,Y_f)$;
    \State $SAT_{circut} = SAT_{circut} \wedge DIVC$;
\EndWhile

\State $KeyGenCircuit = DIVC \wedge (K_1 = K_2)$
\State $Key \leftarrow SAT_F (KeyGenCircuit)$

\end{algorithmic}
\end{algorithm}
\vspace{-4pt}

This reevaluation redirected the attention of the researchers to find harder obfuscation schemes that are more resilient to SAT attacks. SARLock and Anti-SAT \cite{7495588}\cite{xie2016mitigating} obfuscation methods were proposed for this purpose, however further research proved that these obfuscation techniques are prone to a simple removal attack after identification of these blocks using Signal Probability Skew (SPS) attack \cite{7858346}, or identification of most key values using approximate SAT attacks \cite{7951805}, leaving the problem of finding a SAT and SPS resilient obfuscation still unresolved.

A different direction for obfuscating a netlist was proposed in \cite{Shamsi:2017:COC:3060403.3060458} where by introducing feedbacks in the netlist, the netlist is no longer a Directed Acyclic Graph (DAG). In their approach each intentionally created cycle had more than one way to be opened, making such cycle irreducible by structural analysis, claiming that the existence of such cycle breaks the original SAT attack in \cite{7140252}\cite{el2015integrated}. This cyclic obfuscation was later broken with the introduction of CycSAT attack in \cite{8203759}. In CycSAT attack, before invoking the SAT solver, the netlist is checked for key conditions that may result in the creation of cycles. This conditions are translated to a set of cycle avoidance clauses and are added to the list of clauses that represent the circuit SAT problem. The algorithm \ref{CycSAT_algoritm} illustrates the flow of utilizing the cycle avoidance-clauses in CycSAT attack.

\begin{algorithm}
\caption{CycSAT Attack on Cyclic Obfuscated Circuits \label{CycSAT_algoritm}}
\begin{algorithmic}[1]
\scriptsize
\State Find a set of feedback signals $(w_0, w_1, ...w_m)$;
\State Compute "no structural path" formulas $F(w_0, w'_0)$, ..., $F(w_m, w'_m)$;
\State $NC(K)=\wedge^m_{i=0}F(w_i,w'_i)$
\State $C(X,K,Y) = C(X,K,Y)\wedge NC(K)$
\State $SAT_{circut} = C(X,K_1,Y_1) \wedge C(X,K_2,Y_2) \wedge (Y_1 \ne Y_2)$;
\While {$((X_{DI},K_1,K_2)\leftarrow SAT_F(SAT_{circut})=T)$} 
    \State $Y_f \leftarrow C_{BlackBox}(X_{DI})$; 
    \State $DIVC = DIVC \wedge C(X_{DI},K_1,Y_f) \wedge C(X_{DI},K_2,Y_f)$;
    \State $SAT_{circut} = SAT_{circut} \wedge DIVC$;
\EndWhile

\State $KeyGenCircuit = DIVC \wedge (K_1 = K_2)$
\State $Key \leftarrow SAT_F (KeyGenCircuit)$

\vspace{-1pt}
\end{algorithmic}
\end{algorithm}

In this algorithm $(w_0, w_1, ...w_m)$ is a collection of feedback signals whose break will make the encrypted circuit acyclic and $w'_i$ is a signal that feeds to $w_i$ before the break. The function $F(w_i,j)$ is a function that construct the condition for "\emph{having no structural path}" between signal $w_i$ to signal $j$. The  $F(w_i,j)$ is computed by starting from a feedback signal $w_i$ and constructs a string of clauses that satisfy the following condition while traversing a cycle:
\vspace{-3pt}
\begin{equation} \label{eq:erl}
F(w_i,j) = \bigwedge_{l\in NK(j)}F(w_i,l)\lor bk(l,j)
\vspace{-3pt}
\end{equation}

In this function, the $NK(j)$ are the non-key inputs of signal j, and $bk(l,j)$ is the condition on key assuring key does not affect j. This function is initiated with condition $F(w_i,w_i)=0$ and finishes after completing the loop. In this case, the condition for no structural path is tested on all discovered feedback signals in line 3 of the algorithm.


\section{Breaking CycSAT}\label{breakingCycSAt}






The computation for $F(w_i,j)$ could be done in two ways: (i) traversing through a cycle starting from $w_i$ until $w_i$ is visited again and ignoring the cycle break conditions imposed by fanins of other nested cycles; or (ii) traversing through one cycle and adding the cycle break conditions imposed by other nested cycle. We demonstrate that the first choice results in missing some NC conditions, leaving cycles in a design that could break the SAT solver, and by choosing the condition (2) we are forced to build the NC condition by visiting all cycles in the netlist.

\begin{figure}[t]
    \centering

    \begin{subfigure}[b]{70pt}
    \includegraphics[width=0.9\columnwidth]{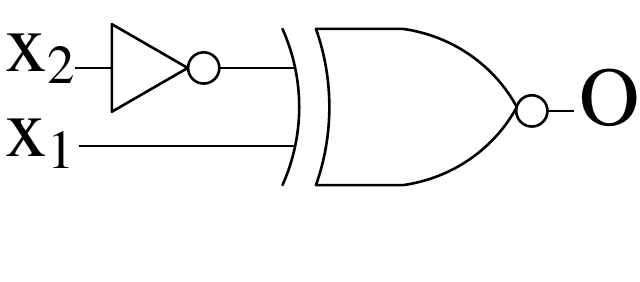}
    \vspace{-12pt} 
    \caption{}

    \vspace{1ex}
    
    \includegraphics[width=0.9\columnwidth]{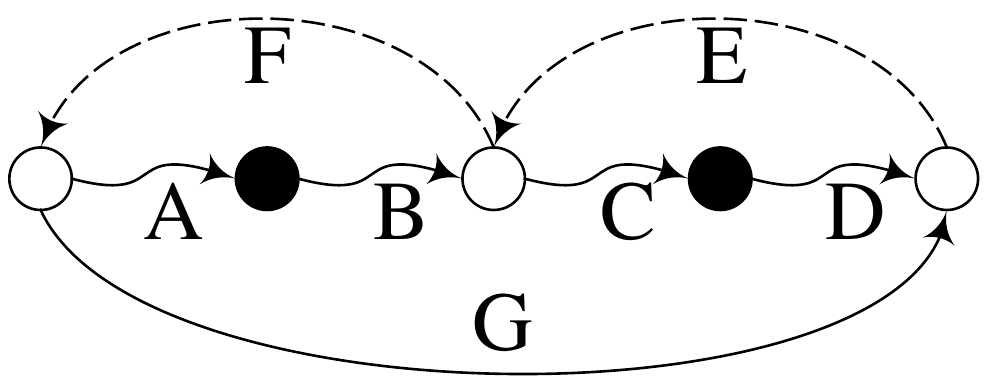}
    \caption{}    
    \end{subfigure}\qquad
    \begin{subfigure}[b]{150pt}
    \includegraphics[width=0.9\columnwidth]{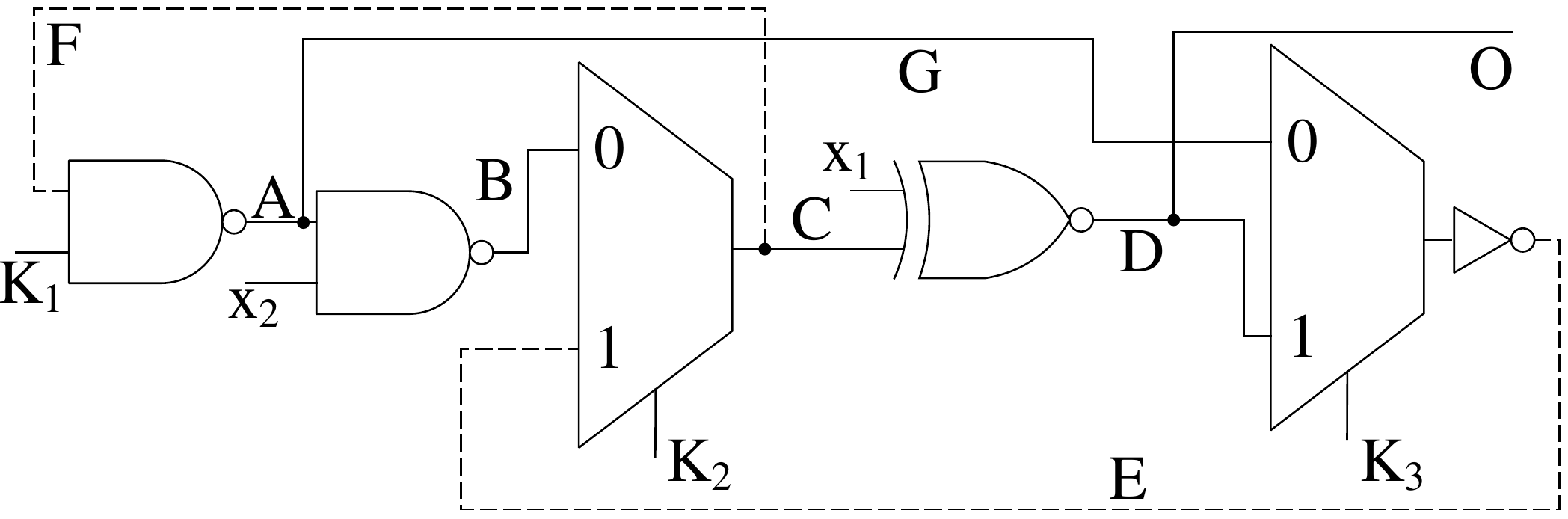}
    \caption{}
    \vspace{10pt}
    \end{subfigure}
    
    \caption{(a) Original circuit (b) Flow diagram of the netlist (c) Obfuscated circuit.}
    \label{graphloop}
\end{figure}


Considering the obfuscated netlist in Fig. \ref{graphloop} and a topological sort from gate A, the edge $E$ and $F$ are identified as feedbacks. When following rule (i), and building the NC condition we will have:

\begin{algorithmic}[1]
\scriptsize
\State $F(F,A) = F(F,F) \lor bk(k_1) = k_1'$ 
\State $F(F,F') = F(F,A) \lor bk(k_2) = k_1' \lor k_2$
\State $F(E,C) = F(E,E) \lor bk(k_2) = k_2'$
\State $F(E,E') = F(E,C) \lor bk(k_3) = k_2' \lor k_3'$
\State $NC=F(F,F')\wedge F(E,E') = (k_1' \lor k_2) \wedge (k_2' \lor k_3')$
\end{algorithmic}

The problem with this assignment is when $(k_1,k_2,k_3)=(0,1,0)$. In this case, the NC condition is satisfied, however, the larger nested cycle $EFGE$ is not broken. Hence, the NC condition would not resolve the cycles if nested or multi-path scenarios exist. In this case, if the wrong key $(k_1,k_2,k_3)=(0,1,0)$ is chosen by SAT solver, it will enter a loop. Depending on whether the cycle is stateful or oscillating, the SAT solver will either be trapped in an infinite loop or will exit with an incorrect key assignment. Note that this infinite loop happens during the execution of the SAT solver and not during the topological sort used in the original SAT attack proposed in \cite{7140252}\cite{el2015integrated}.

To avoid the problem imposed by rule (i), we need to follow the rule (ii) where the key contribution of all fanins in all stages are considered. When using rule (ii) for building the NC condition for the same circuit we have:

\begin{algorithmic}[1]
\scriptsize
\State $F(F,A) = F(F,F) \lor bk(k_1) = k_1'$ 
\State $F(F,F') = (F(F,A) \lor bk(k_2)) \wedge (F(F,E) \lor bk(k_2)) = (k_1' \lor k_2) \wedge (k_1' \lor k_3 \lor k_2'$) 
\State $F(E,C) = F(E,E) \lor bk(k_2) = k_2'$
\State $F(E,E') = (F(E,C) \lor bk(k_3)) \wedge  (F(E,G) \lor bk(k_3))  = (k_2' \lor k_3') \wedge (k_2' \lor k_1' \lor k_3)$
\State $NC=F(C,C')\wedge F(E,E') = (k_2' \lor k_3') \wedge (k_1' \lor k_2' \lor k_3) \wedge (k_1' \lor k_2).$
\end{algorithmic}
\textcolor{white}{.}
By following rule (ii), the previous assignment of keys $(k_1,k_2,k_3)=(0,1,0)$ will no longer be a valid assignment, preventing the SAT solver from being stuck or suggesting a wrong key. However, in this case, \emph{all cycles in the design have to be traversed and conditioned}. As a matter of fact, given the way the NC is formulated in \cite{8203759}, in order to derive the "no structural path" condition, some of the combinational cycles (such as $EFGE$ in Fig. \ref{graphloop}) have been visited more than once. Hence, the number of times the key conditions has to be generated is even larger than the number of cycles in a netlist.



The problem of visiting nested cycles more than once in a CycSAT attack could be resolved by a slight modification to the CycSAT pre-processing step. In the modified attack, instead of applying rule (ii) on one-cycle-per feedback, we could apply the rule (i) on all cycles. It is intuitive to see that both approaches produce the same NC clauses. For example, in Fig. \ref{graphloop} when following condition (i), and traversing cycle $EFGE$, the condition $(k_1' \lor k_2' \lor k_3)$ is generated. Hence, by ANDing the generated condition to the two clauses generated by applying the rule (i), the NC condition of rule (ii) is generated. However, in this case, the combinational cycle $EFGE$ is only visited once. Even by considering the improvement suggested in CycSAT formulation, it still requires visiting all cycles in a netlist to compose the NC clauses. This necessity is used to break the CycSAT attack in this paper.

A different method of introducing complexity is by eliminating DAG nature of the original netlist and by transforming it into a Boolean cyclic function, which could be represented using a Directed Cyclic Graph (DCG), before subjecting it to cyclic obfuscation. If the original netlist is not a DAG, the CycSAT pre-processing step has to build the NC condition by checking for "\emph{no sensitizable path}" condition \cite{8203759}, instead of "\emph{no structural path}" condition.  The no sensitizable path condition from \cite{8203759} is recited in equation \ref{eq:NSenP}: 
\vspace{-4pt}
\begin{equation} \label{eq:NSenP}
F(w_i,j) = \bigwedge_{l\in fanin(j)}F(w_i,l)\lor ns(l,j)
\end{equation}
\vspace{-9pt}

The "no sensitizable path" condition generates a  clause for each multi-input gate in a cycle. As the result, NC clauses are much longer and much weaker. Hence, adding even a small number of feedbacks to such circuits (that have valid Boolean cycles) for the purpose of obfuscation, will significantly increase the size of the circuitSAT problem. To illustrate the weaker and longer nature of the NC clauses, the no "sensitizable path condition" for the circuit in Fig. \ref{graphloop} is constructed below:

\begin{algorithmic}[1]
\scriptsize
\State $F(F,A) = F(F,F) \lor ns(F,A) =\enspace k_1'$ 
\State $F(F,B) = F(F,A) \lor ns(A,B) = k_1' \lor x_2'$
\State $F(F,F') = (F(F,B) \lor ns(B,F')) \wedge (F(F,E) \lor ns(E,F')) = (k_1' \lor x_2' \lor k_2) \wedge (k_1' \lor k_3 \lor k_2'$) 
\State $F(E,C) = F(E,E) \lor ns(E,C) = k_2'$
\State $F(E,D) = F(E,C) \lor ns(C,D) = k_2'$
\State $F(E,E') = (F(E,D) \lor ns(D,E')) \wedge (F(E,G) \lor ns(G,E')) = (k_2' \lor k_3') \wedge (k_2' \lor k_1' \lor x_2' \lor k_3)$
\State $NC=F(F,F')\wedge F(E,E') = (k_1' \lor x_2' \lor k_2) \wedge (k_1' \lor k_3 \lor k_2') \wedge (k_2' \lor k_3') \wedge (k_2' \lor k_1' \lor x_2' \lor k_3)$
\end{algorithmic}

The CycSAT pre-processing time, as illustrated in equation \ref{eq:preproceTime}, is linearly related to the number of discovered cycles $N$ and the time for composing the NC condition per cycle $t_{NC}$. Our approach for breaking the CycSAT is to exponentially increase the time needed for composing the NC condition in the pre-processing step of CycSAT beyond acceptable. This is achieved by exponentially increasing the number of cycles $N$ in a design with respect to the number of inserted feedbacks $m$, and increasing the time required for processing each cycle ($t_{NC}$) by forcing the pre-processing step to consider the "no sensitizable path" condition instead of " no structural path" condition.
\vspace{-10pt}
\begin{equation} \label{eq:preproceTime}
T_{NC} = \sum_{i=1}^N t_{NC} \enspace | \enspace  N = Ae^{m}
\end{equation}

In the next section we propose two methods for building an exponential relation between the number of cycles in a netlist with the number of inserted feedbacks and subsequently introduce three techniques for transforming a netlist to contain cyclic Boolean functions which forces an attacker to use the "no sensitizable path" condition in CycSAT attack. 


\vspace{-6pt}
\section{Cyclic Obfuscation}\label{cyclock}
\subsection{Exponentially increasing the number of cycles in a netlist}
In order to exponentially increase the number of cycles in a given netlist with respect to the number of inserted feedbacks, we introduce two approaches: (1) building Super Cycles (SC) and, (2) building Logarithmic Feedback Networks (LFN). 

\vspace{-4pt}
\subsubsection{\textbf{Building Super Cycles (SC)}}
\begin{figure}[h]
    \centering 

    \begin{subfigure}[b]{188pt}
    \includegraphics[width=\columnwidth]{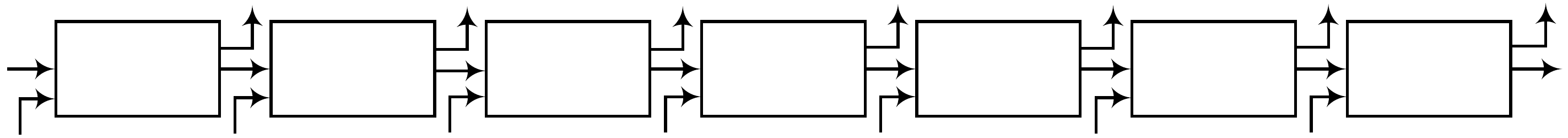}
    \vspace{-17pt}
    \caption{}
    \end{subfigure}

    \begin{subfigure}[b]{220pt}
    \includegraphics[width=\columnwidth]{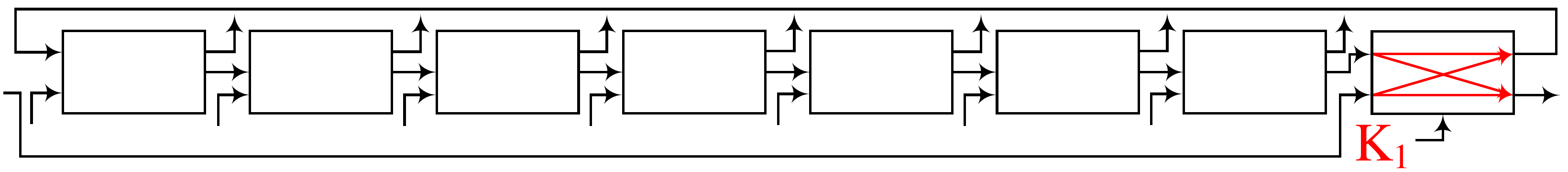}
    \vspace{-17pt}
    \caption{}
    \end{subfigure}

    \begin{subfigure}[b]{220pt}
    \includegraphics[width=\columnwidth]{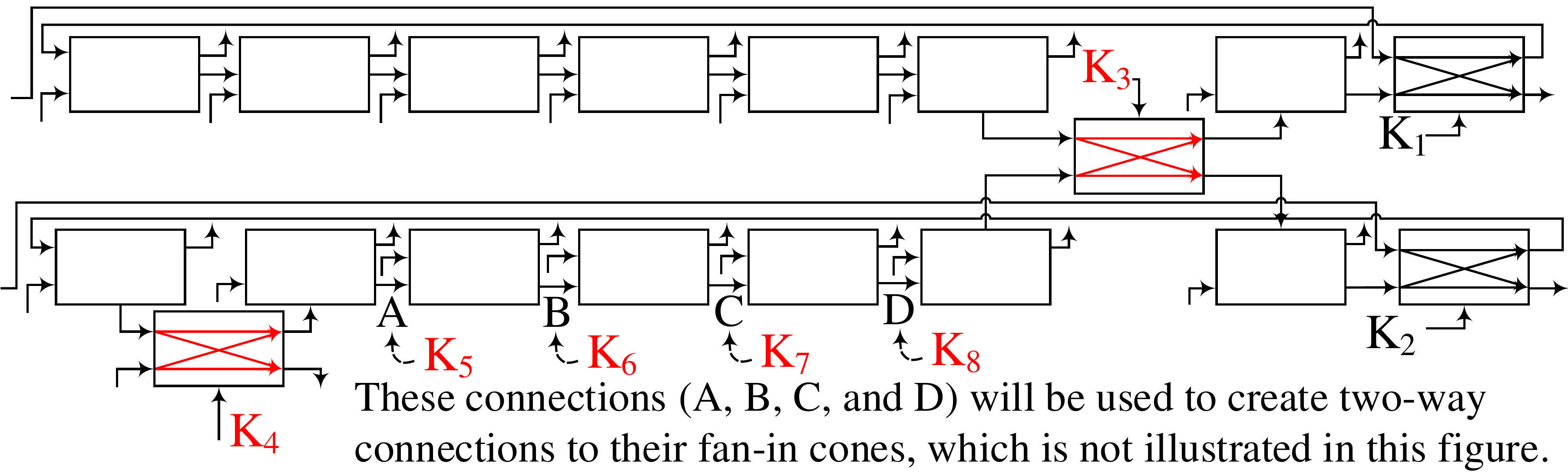}
    \vspace{-17pt}
    \caption{}
    \end{subfigure}

    \caption{Building a Super Cycle from 7 gate MC. (a) A path segment containing 7 gates, (b) building a Micro Cycle, (c) building a SC by strongly connecting multiple MCs.}
    \label{buildingSCfig}
\end{figure}

The process of building a SC is illustrated in Fig. \ref{buildingSCfig}. To define and build a SC, let us first define a Micro Cycle (MC). A MC is a cycle created by following the cycle creation conditions adopted from \cite{Shamsi:2017:COC:3060403.3060458}, which are recited below:\\\\[-7pt]
MC Condition 1: Any created cycle has to be non-reducible\\
MC Condition 2: At least $n\geq2$ edges in each small cycle have to be removable\\[-9pt]

A reducible cycle has a single entry point. Hence, the depth-first-search (DFS) traversal of a netlist, that only contain reducible cycles is unique allowing reducible cycles to be opened by removing a unique set of edges which can be found efficiently \cite{Shamsi:2017:COC:3060403.3060458}. By having multiple entries into each MC, the non-reducable condition is satisfied.  Combined with having more than one removable edge, this forces an adversary to use the CycSAT pre-processing step to generate the necessary cycle avoidance clauses before invoking the SAT solver. At this point, when each MC is considered as a graph vertice, a SC is the strongly connected graph of these vertices. In graph theory, a strongly connected graph is defined as a graph with at least one path between any two pairs of its vertices. Finally in the last step, the edge density of the SC is increased, creating additional paths between MCs. The process of building a SC is captured in algorithm \ref{buildingSC}. 


%

\begin{algorithm}
\caption{Steps for building a Super Cycle \label{buildingSC}}
\begin{algorithmic}[1]
\scriptsize
\State Construct MCs in the fanin of smallest possible number of primary outputs.
\State Strongly connect all MC cycles (this is illustrated in Fig. \ref{buildingSCfig}.b is done by creating a two-way connection between each newly created MC, and the existing SC).
\State Select signals in MCs (A, B, C, D in Fig. \ref{buildingSCfig}.c) that are not used for SC connectivity and provide a two way path from them to unused edges in other MCs or random signals in their fanin cone.
\end{algorithmic}
\end{algorithm}

By forcing the MC cycles to the fanin of the smallest number of primary outputs, we increase the possibility of shared edges or connecting edges between created MCs. By having all MC cycles strongly connected, we create the possibility of larger combinational cycles. And finally, adding the random connections increase the density of the edges in the strongly connected graph, increasing the number of resulting cycles. In the result section, we illustrate that the number of created cycles, by following the steps in algorithm \ref{buildingSC} has an exponential relation with the number of inserted feedbacks.



\vspace{-6pt}
\subsubsection{\textbf{Building Logarithmic Feedback Networks (LFN)}}
\begin{figure}[h]
    \centering
    \begin{subfigure}[b]{160pt}
    \includegraphics[width=\columnwidth]{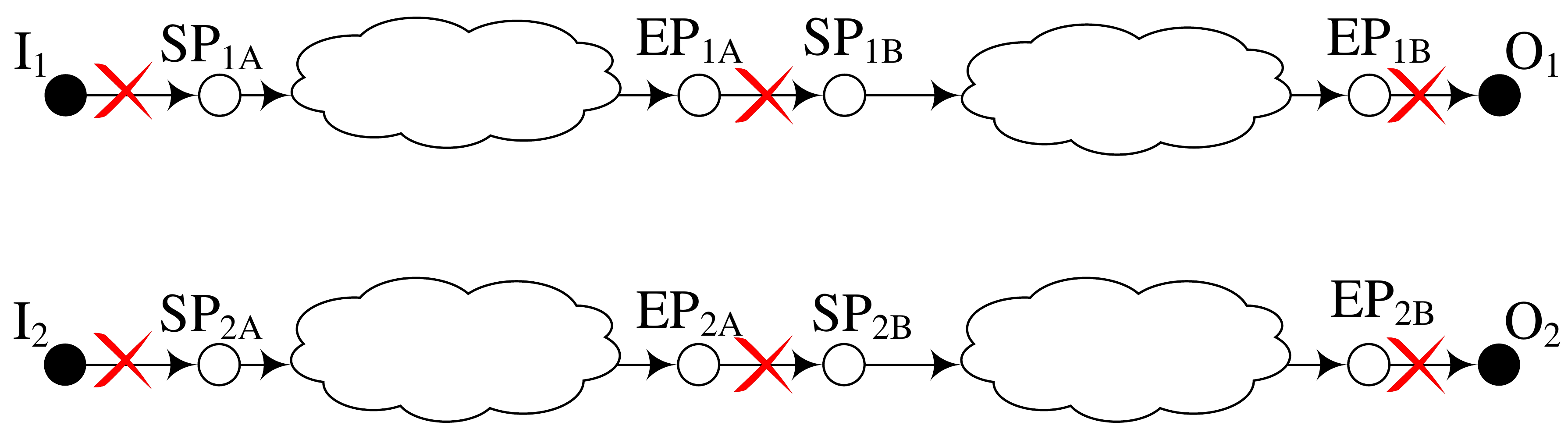}
    \vspace{-17pt}
    \caption{}
    \end{subfigure}
    
    \begin{subfigure}[b]{80pt}
    \includegraphics[width=\columnwidth]{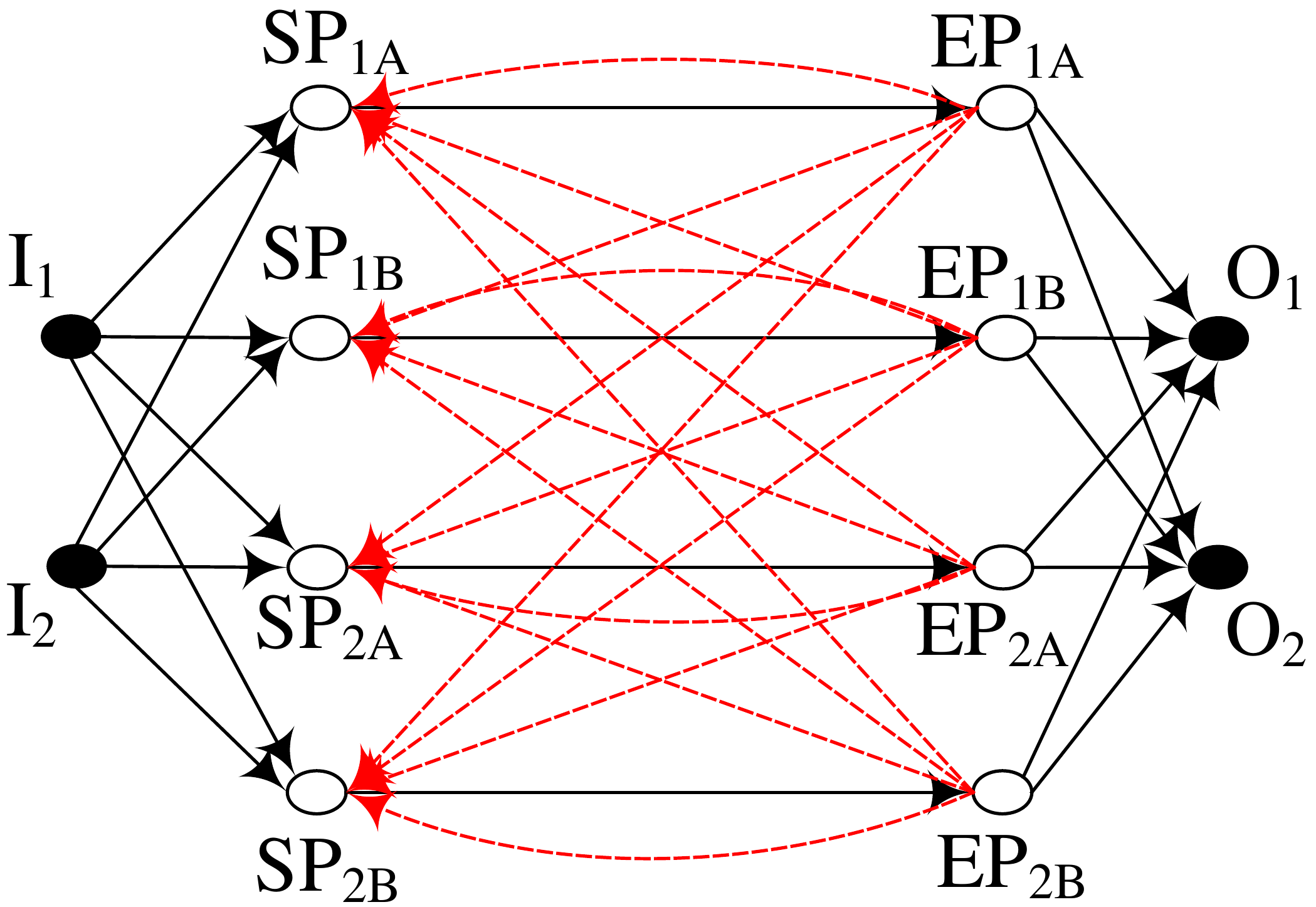}
    \vspace{-15pt} 
    \caption{}
    \end{subfigure}
    \hspace{35pt}
    \begin{subfigure}[b]{80pt}
    \includegraphics[width=\columnwidth]{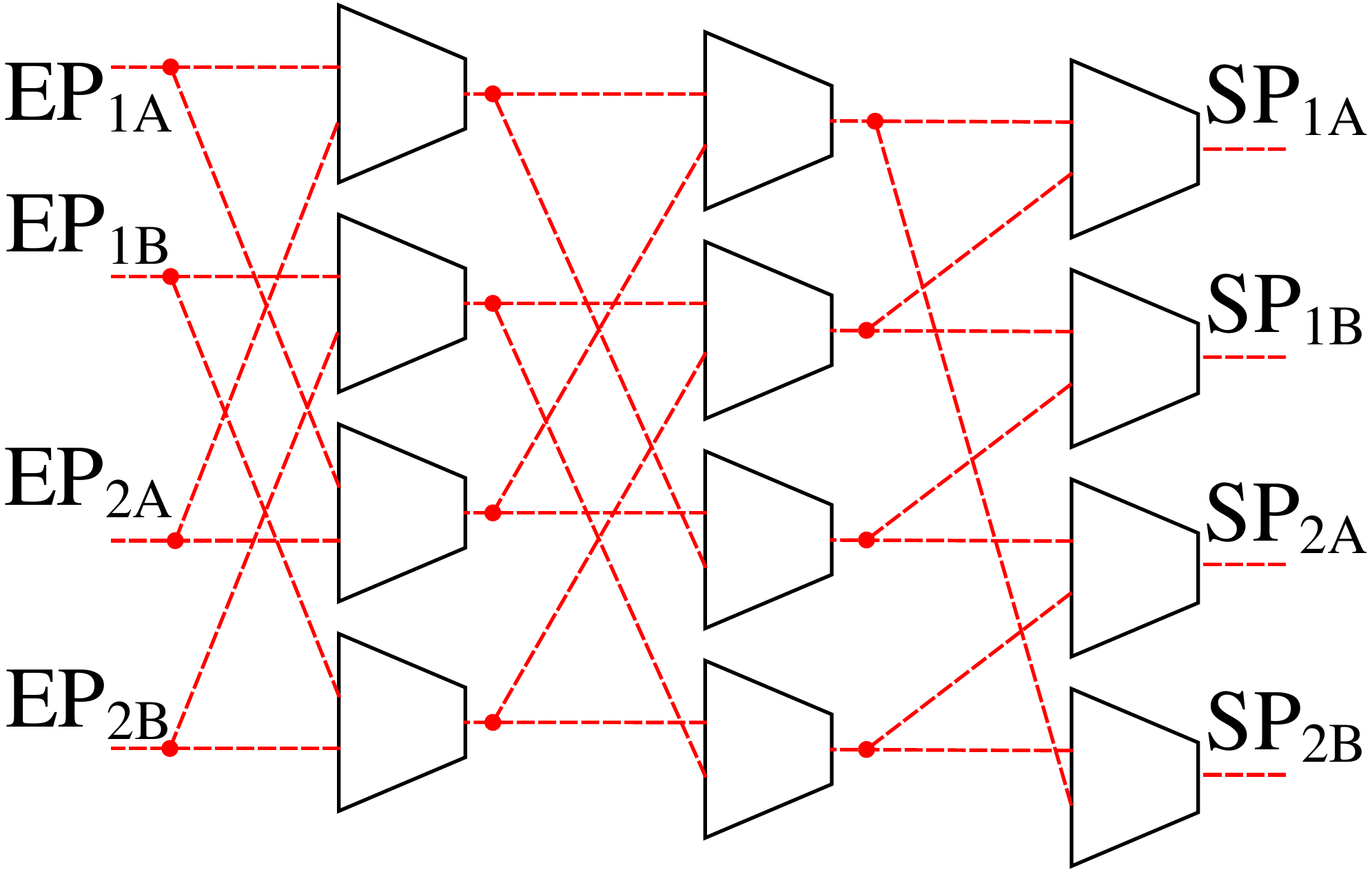}
    \vspace{-15pt} 
    \caption{}
    \end{subfigure}    
    
    \caption{Building a logarithmic feedback network in which the number of cycles exponentially increase with the number of feedbacks.}
    \label{logNet}
\end{figure}

In this method, as illustrated in Fig. \ref{logNet}.a, several path segments in the fanin cone of the same primary output are selected. By breaking a signal in the midpoint of each path, two path segments are created. The signal entering and the signal exiting each half segment is marked as its start point (SP) and end point (EP) respectively. Then the SP and EP of multiple such path segments are used to build a logarithmic switching network (e.g., Omega, Butterfly, Benes, or Banyan network). When connecting $M$ EPs to $M$ SPs we need  $M(1+log_2(M))$ muxes for a non-blocking logarithmic network. In this case, when the correct key is applied, the switching network is configured correctly, otherwise, an invalid connectivity obfuscates the netlist functionality.

\textbf{Lemma.} \emph{The lower bound on the number of cycles created when using LFN is $\sum_{l=1}^m {m \choose l}(l - 1)!$, when $m$ is the number of inserted feedbacks and $l$ is the log base two of the number of cycles of size $l$.}

\emph{Proof.} The proposed LFN is a special case of a complete bipartite graph that contains no odd cycles. Suppose that $SE_{ij}$ indicates a vertex from $SP_i$ to $EP_j$. Similarly, $ES_{ij}$ indicates a vertex from $EP_i$ to $SP_j$. For $l = 2$, the cycles are all paths from a $SP$ to its corresponding $EP$ and return path $\{SE_{ii}, ES_{ii}\}$. If we start from $SP_{i}$, the second visited node is its $EP$ ($EP_i$). Since each $EP$ is connected to all $SP_s$, for intermediate nodes, we have all permutations as possible paths. cycles with $l = 2$, have no intermediate node. So, there are ${m \choose 1}0!$ cycles when $l = 2$. For $l = 4$, the cycles are paths like $\{SE_{ii}, ES_{ij}, SE_{jj}, ES_{ji}\}$. There is only one intermediate node in cycles when $l = 4$ resulting in ${m \choose 2}1!$ cycles. Similarly, for $l = 8$, the cycles are paths like $\{SE_{ii}, ES_{ij}, SE_{jj}, ES_{jk}, SE_{kk}, ES_{ki}\}$. Since, we have two intermediate node, $j$ and $k$, we should consider their permutation as a new cycle, i.e. $\{SE_{ii}, ES_{ik}, SE_{kk}, ES_{kj}, SE_{jj}, ES_{ji}\}$. So, for $l = 8$, we have ${m \choose 3}2!$. We can extend this relation to all cycles with different length. The summation of these cycles indicates the number of cycles in our logarithmic network, which is $\sum_{l=1}^m {m \choose l}(l - 1)!$. $\blacksquare$

Note that the $\sum_{l=1}^m {m \choose l}(l - 1)!$ is the lower bound of the number of simple and nested cycles created by using the logarithmic network. As a matter of fact, the number of paths from each SP to each EP could be more than 1, and there are possibilities of having a connection between SPs and EPs of the different paths in the original circuit, increasing the number of cycle possibilities to a far larger number. Based on the lower bound formula, the number of created cycles is $O(\sum_{l=1}^m {m \choose l}(l - 1)!) \leq  O(m!) = O(m^m)$. Hence, there exists an exponential relation between the number of inserted feedbacks and the number of resulting cycles in the netlist. \vspace{-10pt}

\subsection{Building Cyclic Boolean Functions}
A Boolean function does not need to be acyclic; it is possible to reduce the number of gates in a circuit if a function could be implemented in its acyclic form \cite{7406959}\cite{1466160}\cite{1218927}\cite{Rivest:1977:NFM:1310165.1310794}. For example, the work in \cite{Rivest:1977:NFM:1310165.1310794} presents an n-input 2n-output positive unate Boolean function which can be realized with $2n$ two-input gates when feedback is used but requires $3n-2$ gates if feedback is not used. Hence, cyclification of a circuit in addition to forcing the CycSAT pre-processing step to consider the "no sensitizable path", could also remedy the area overhead of introducing new gates for cyclic obfuscation. To cyclify a netlist and to increase the $t_{NC}$ in equation \ref{eq:preproceTime}, we suggest three approaches: (1) Template-based cyclic-function mapping, (2)  Input-dependency based cycle generation  and, (3) Node-merging based cycle generation.

\begin{figure}[h]
\centering
    \includegraphics[width=0.6\columnwidth]{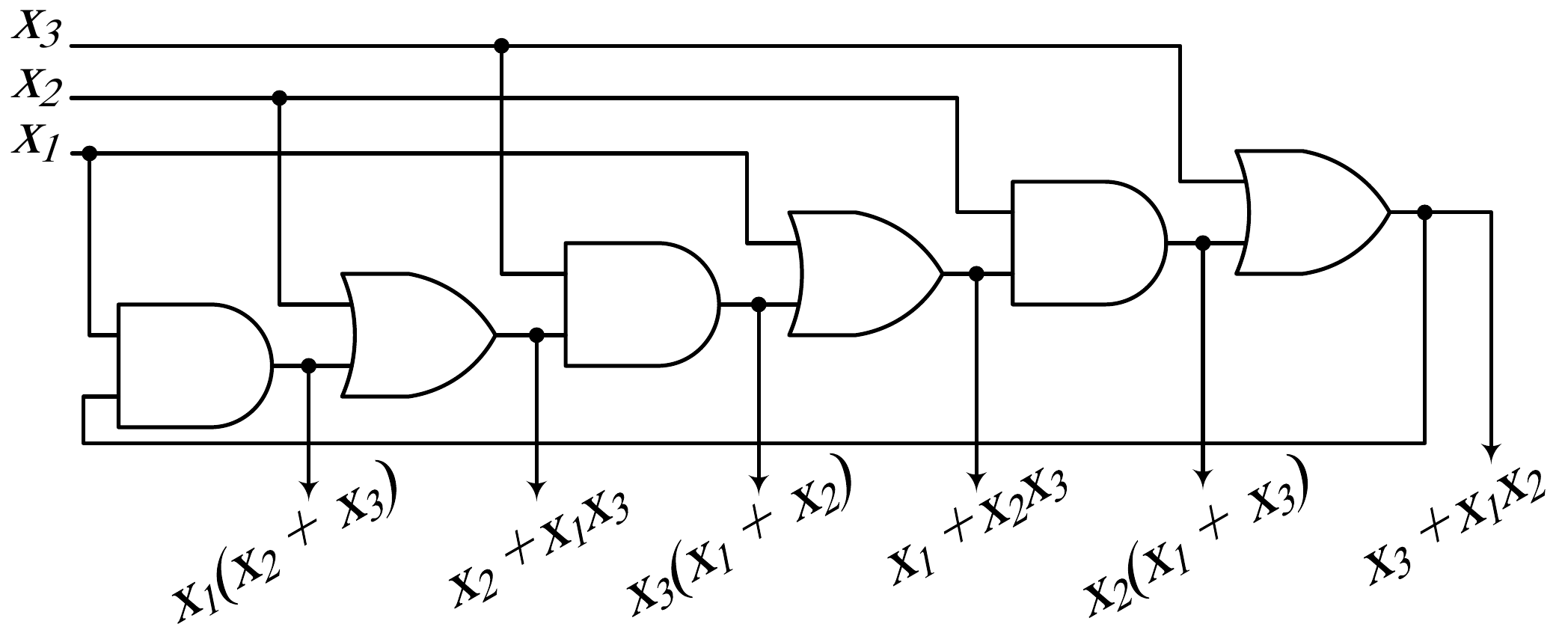}
    \caption{3-input Rivest circuit implementing six functions.}
    \label{Rivest}
    \vspace{-5pt}
\end{figure}

\vspace{-3pt}
\subsubsection{\textbf{Template-based cyclic-function mapping}}
In this approach, many small cyclic Boolean circuits are collected as templates in our obfuscation library. Then, A netlist is scanned for opportunities (with and without logic manipulation) to replace a cluster of logic gates with such templates. An example of such feedback template is the circuit introduced in \cite{Rivest:1977:NFM:1310165.1310794} where a special case of it (for 3 inputs) is illustrated in Fig. \ref{Rivest}. To introduce cycles, the circuit could be modified to introduce at least one of the possible functions in this circuit. The candidate logic cluster is then replaced by the template. To prevent template scanning and removal attacks, in a subsequent camouflaging step (by means of gate and route obfuscation) the template will be hidden. Note that many such templates could be made \cite{Rivest:1977:NFM:1310165.1310794}\cite{7406959}\cite{1466160}\cite{1218927}, and by not knowing the template type and the camouflaged technique used to hide the connection, an attacker has no prior information to identify and remove these templates.   

\begin{figure}[h]
\centering
    \includegraphics[width=0.70\columnwidth]{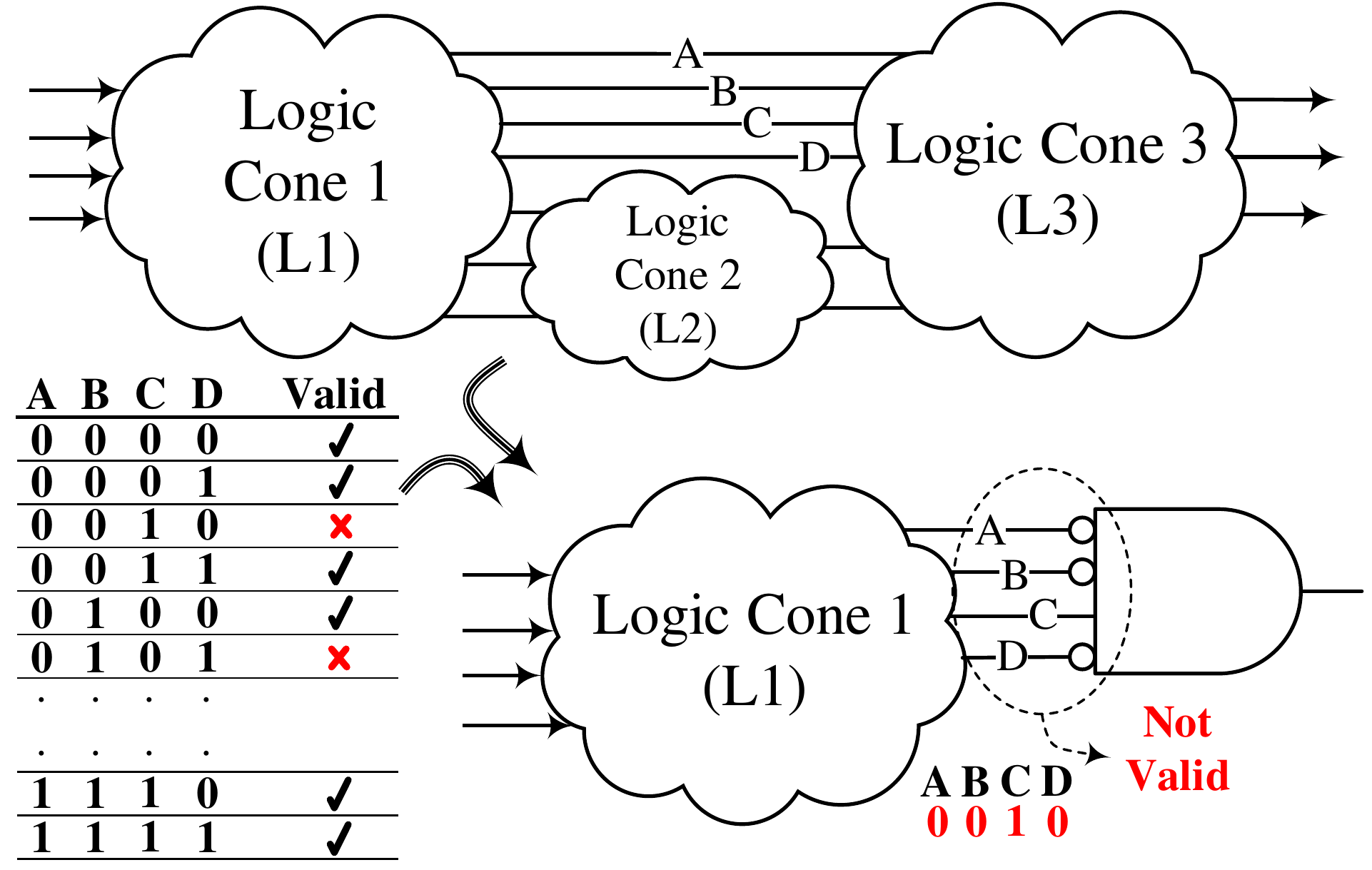}
    \caption{Due to correlation of intermediate signals, certain signal combinations may never occur.}
    \label{SigDepend}
\end{figure}

\vspace{-3pt}
\subsubsection{\textbf{Input-dependency based cycle generation}} \label{idbo}
This method explores the correlations between signals that share common primary inputs in their fanin cone. Considering N such signals in an arbitrary stage of a DAG, some of the $2^N$ input possibilities may never occur. For example, when tracking 4 signals $A$, $B$, $C$, and $D$ in Fig. \ref{SigDepend}, we may find that $ABCD=\{0010\}$ could not occur. A SAT solver cold be used for finding the non-occurring input scenarios; This process is  illustrated in Fig. \ref{SigDepend}, where the logic clusters L2 and L3 are removed, and the 4 signals are ANDed together such that for a certain case, for example, $ABCD=0010$, the output of AND gate is evaluated to 1. Then, this circuit is given to a SAT solver to find a satisfying input assignment. If SAT solver returns UNSAT, this combination of input is chosen since it would never happen, otherwise, a different combination is checked.

\begin{figure*}[t]
\centering
    \begin{subfigure}[b]{92pt}
    \includegraphics[width=\columnwidth]{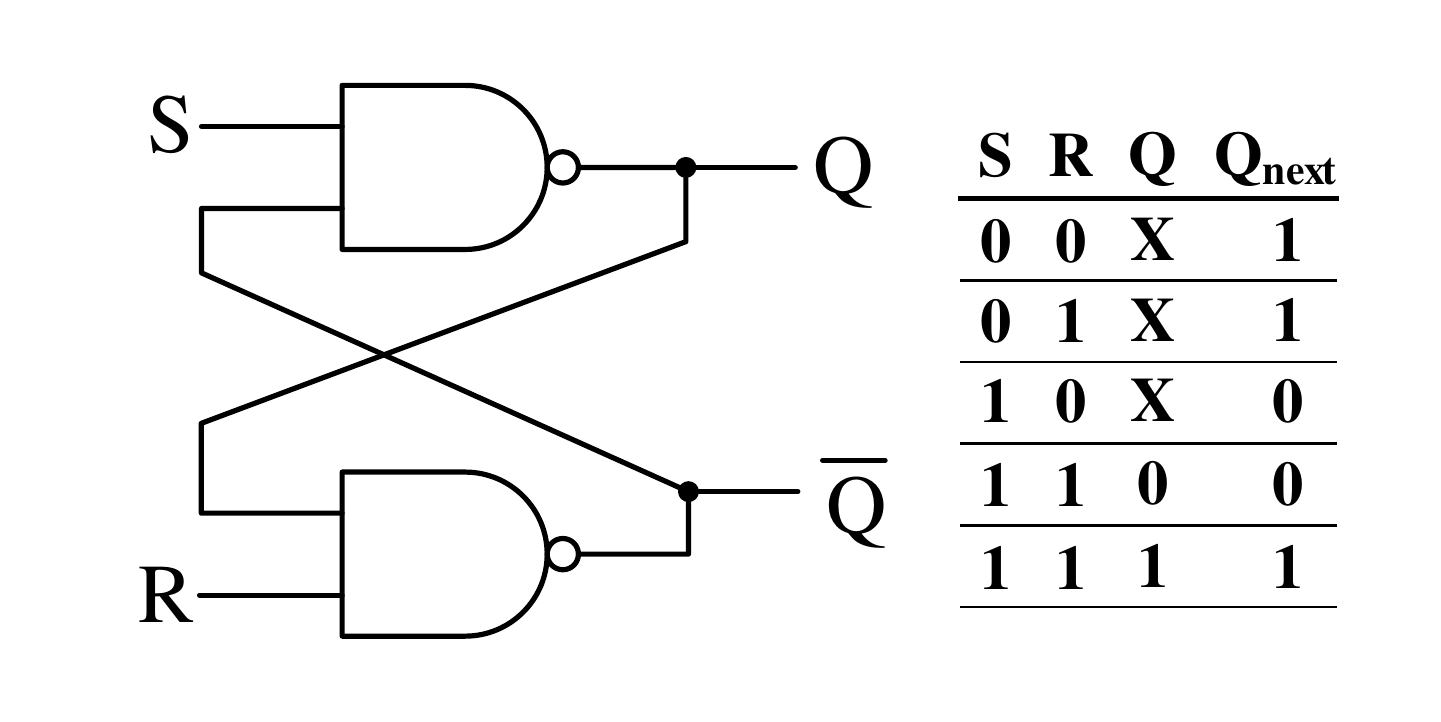}
    \vspace{-17pt}
    \caption{}
    \end{subfigure}
    \hspace{10pt}
    \begin{subfigure}[b]{92pt}
    \includegraphics[width=\columnwidth]{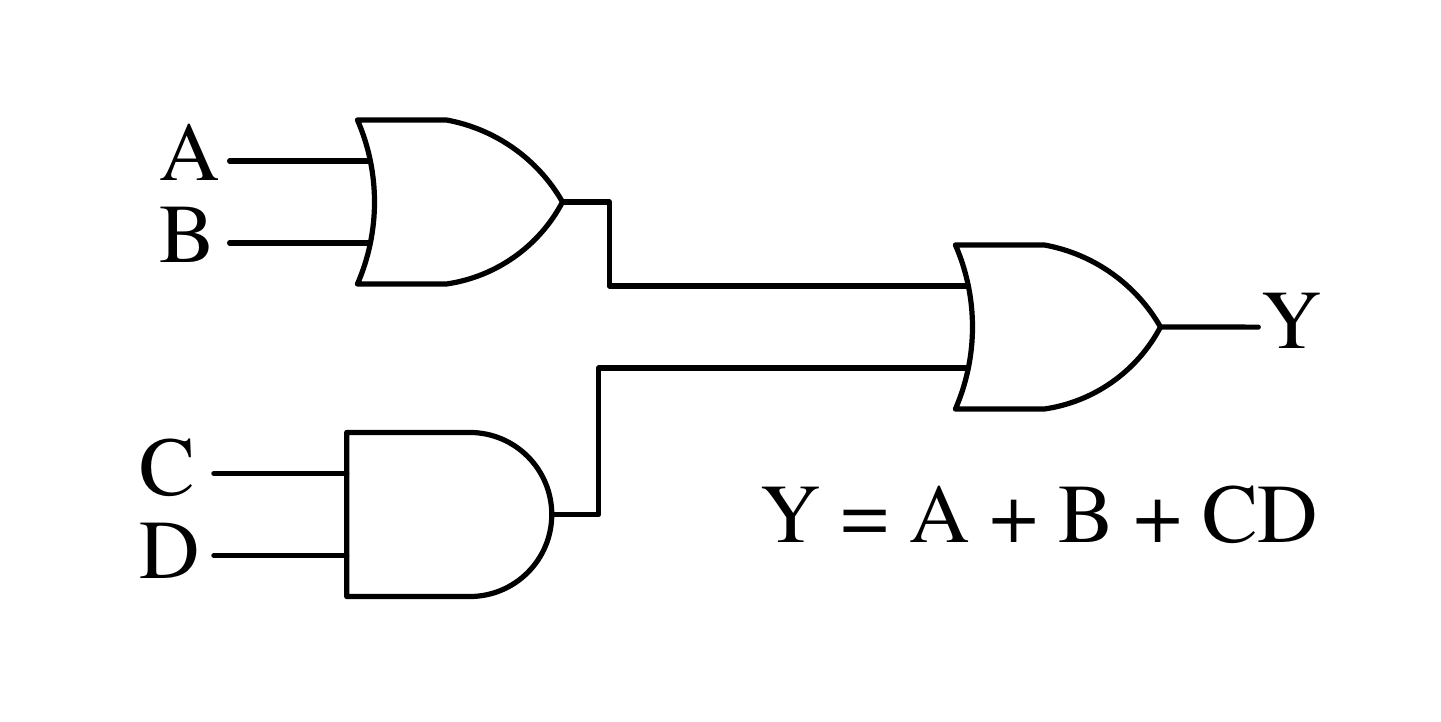}
    \vspace{-17pt}
    \caption{}
    \end{subfigure}
    \hspace{10pt}
    \begin{subfigure}[b]{92pt}
    \includegraphics[width=\columnwidth]{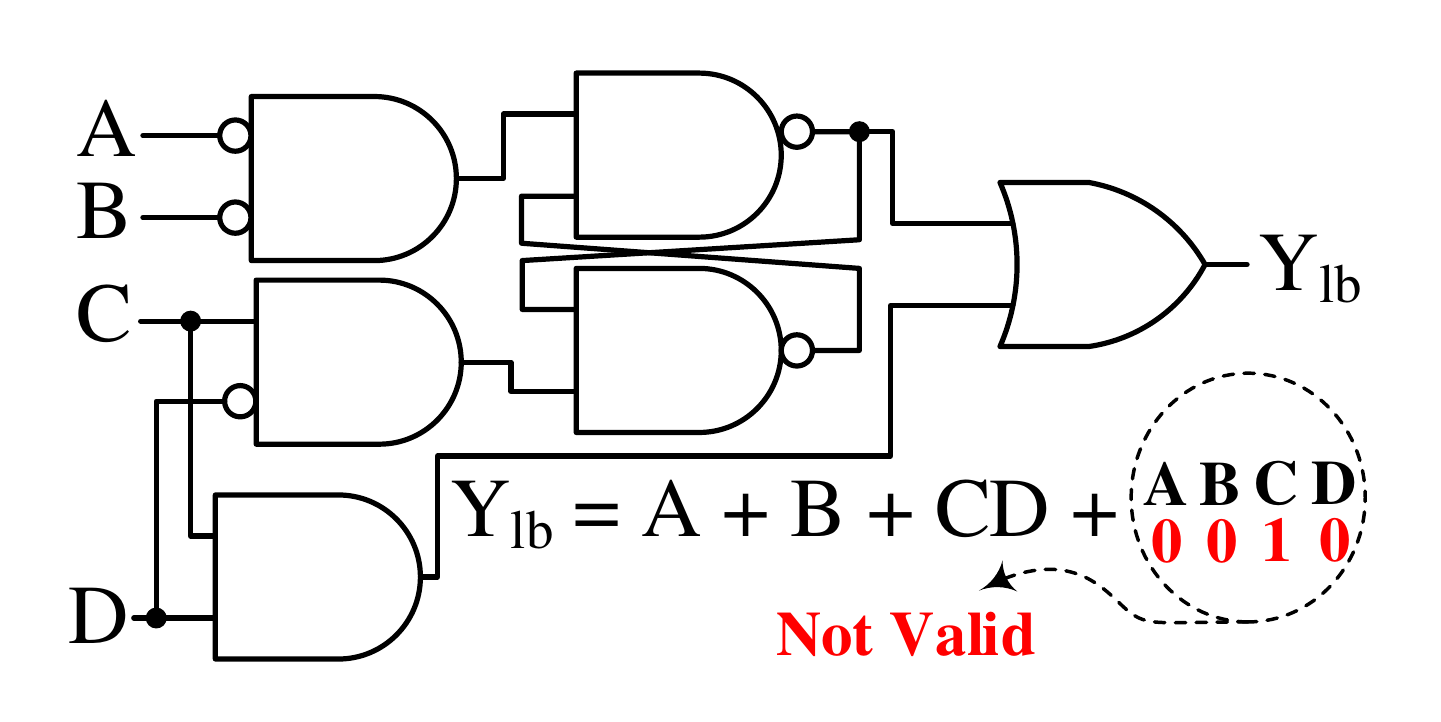}
    \vspace{-17pt}
    \caption{}
    \end{subfigure}
    \hspace{15pt}
    \begin{subfigure}[b]{148pt}
    \includegraphics[width=\columnwidth]{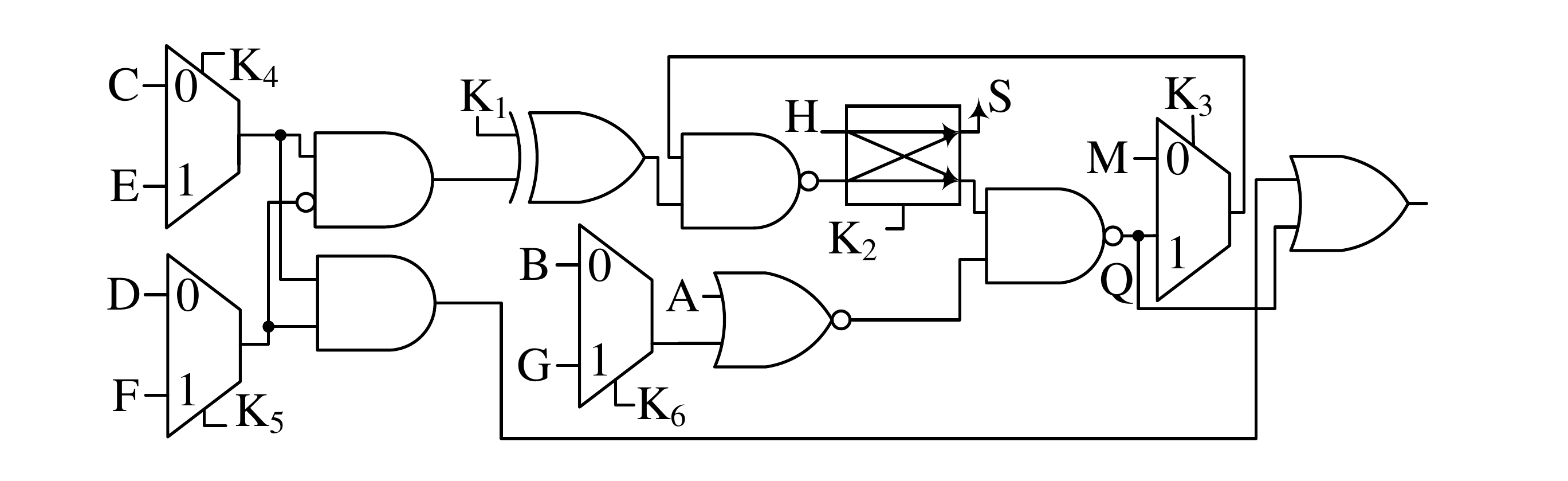}
    \vspace{-17pt}
    \caption{}
    \end{subfigure}

    \caption{Input-dependency-based cyclification of a Boolean function. (a) SR latch (b) original circuit (c) cyclified circut when $ABCD=0010$ is non-occuring. (d) obfuscated cyclified circuit using additonal random inputs $E,F,G,H$ and $M$.}
    \label{latchobf}
\end{figure*}

In the next step, we use a sequential element and tie the discovered non-occurring input scenario to the state preserving input of the sequential element. For example, by using SR latch in Fig. \ref{latchobf}.a, If $SR=11$ doesn't happen, the $Q_{next}$ is the inverse of input $S$. Hence, we can build a circuit that ties the discovered non-occurring input scenario to the $SR=11$. For example, let's assume wires A, B, C and D have a non-occurring combination $ABCD=0010$ and these signals construct the signal $Y=A+B+CD$. Fig \ref{latchobf}.c illustrates the signal $Y$ reconstructed when the non-occurring combination of the inputs is tied to SR input of the latch. In the next step, to hide the correlation between input signals, they are further obfuscated. The SR latch feedback is also obfuscated using a set of muxes that create alternative paths for its feedback signals. This assures that the CycSAT can only generate the correct NC clauses if the "no sensitizable path" condition is processed, otherwise, it breaks the SR latch feedback and invalidates the netlist.

\vspace{-5pt}
\subsubsection{\textbf{Node-merging based cycle generation}}
The third approach for cyclification of a netlist is based on the work in \cite{7406959} where the logic implication is used to identify cyclifiable structure candidates directly, or to create them aggressively in circuits. At its core, the work in \cite{7406959} introduces active combinational feedback cycles by merging two nodes in the original DAG. To check the validity of the generated cyclic netlist, they use a SAT-based algorithm and validate whether the formed cycles are combinational or not.  \vspace{-4pt}

\section{Results}\label{results}
In this section, we analyze the effectiveness of our proposed defense against SAT and CycSAT attacks. For finding the list of cycles in a netlist after cyclic obfuscation, the algorithm in \cite{hawick2008enumerating} was implemented in C++. Our computational platform is a Dell PowerEdge R620 equipped with Intel Xeon E5-2670 2.50GHz and 64GB of RAM.  We have used our proposed Super Cycle scheme for exponentially increasing the number of cycles on ISCAS-85 benchmarks, and the input-dependency-based netlist cyclification for forcing the CycSAT pre-processing to use the "no sensitizable path" condition instead of "no structural path" condition. However, similar results are observed when LFN is deployed, or other proposed techniques (template-based cyclification, and node-merging) are used for cyclification of a netlist.

\begin{table}[h]
\footnotesize 
\centering
\caption{SAT-attack and CycSAT execution time after insertion of a SC (N=2) and insertion of a SC and 10 SR latches (N=2 + SR-L=10).}
\label{sat_results}
\setlength\tabcolsep{2.5pt} 
\begin{tabular}{@{} *8c @{}}
\toprule 
        \multirow{ 2}{*}{Circuit} & \multicolumn{3}{c}{N=2} & \multicolumn{4}{c}{N=2 + SR-L=10} \\ 
        \cmidrule(lr){2-4}
        \cmidrule(lr){5-8}
        & SAT & \#Cycles & CycSAT-I & SAT & \#Cycles & CycSAT-I & CycSAT-II \\ \midrule
c432    & Inf Loop & 23,879 & 2.561 s  & Inf Loop & $1.65 \times 10^5$ & UNSAT & 11.691 s \\ 
c499    & 0.56 s & 236 & 0.104 s  & Inf Loop & 397 & UNSAT & 0.118 s \\
c880    & Inf Loop & 1,601 & 0.245 s  & Inf Loop & $7.87 \times 10^6$ & UNSAT & 793.125 s \\
c1355   & Inf Loop & 636 & 0.122 s  & Inf Loop & $5.00 \times 10^5$ & UNSAT & 53.215 s \\
c1908   & 0.28 s & 294 & 0.101 s  & Inf Loop & 6,467 & UNSAT & 0.732 s \\
c2670   & Inf Loop & 1,570 & 0.234 s  & Inf Loop & 7,412 & UNSAT & 0.927 s \\
c3540   & Inf Loop & 5,991 & 0.756 s  & Inf Loop & 6,026 & UNSAT & 0.753 s \\
c5315   & Inf Loop & 4,869 & 0.613 s  & Inf Loop & $2.59 \times 10^5$ & UNSAT & 26.042 s \\
c7552   & Inf Loop & 124 & 0.189 s  & Inf Loop & 164 & UNSAT & 0.195 s \\ \bottomrule
\end{tabular}
\end{table}

Table \ref{sat_results} captures the behavior of SAT and CycSAT when dealing with obfuscated cyclic and acyclic netlists. For generating the data in this table, we have created a cyclic version of each ISCAS-85 benchmark using input-dependency based obfuscation proposed in section \ref{idbo}, and then implemented a super cycle with two MCs in each netlist. The pure SAT attack is trapped in an infinite loop in both cases, with an exception of two benchmarks, that SAT solver luckily chooses a sequence of keys that avoid or exit the trap. The CycSAT when uses the "no structural path" condition (CycSAT-I) for the acyclic circuit, breaks the obfuscation easily, however, as illustrated in this table and predicted in equation \ref{eq:preproceTime}, its runtime (which include the runtime for both pre-processing step and SAT solver's invocation) almost linearly varies with the number of cycles in each netlist. But when the netlist is acyclic, CycSAT-I returns UNSAT as it produces NC clause that breaks the DCG cycles incorrectly. On the other hand, the CycSAT when uses the "no sensitizable path" condition (CycSAT-II), breaks the obfuscation in all cases. However, since the number of created cycles are larger, and the time it takes to compose the NC condition for each cycle based on "no sensitizable path" condition is longer, the runtime of the SAT solver is considerably longer. In short, this table shows the capabilities of SAT, and CycSAT (I and II) for dealing with cyclic and acyclic original netlists, and also express the linear dependence between the CycSAT runtime and the number of created cycles. Hence, if the number of cycles exponentially increase, the runtime of CycSAT (pre-processing step) also exponentially increase.

\begin{table}[b]
\scriptsize
\centering
\caption{The number of cycles reported during CycSAT attack. The exponential fitting function is in form of $y=Ae^{Bx}$.}
\label{table_of_SC}
\setlength\tabcolsep{1.2pt} 
\setlength\extrarowheight{2pt}
\begin{tabular}{@{} l *9c @{}}
\toprule
\multicolumn{1}{c}{Circuit} & N=1 & N=2 & N=3 & N=5 & N=10 & N=15 & N=20 & Exponential Fit Function \\
\midrule
c432    & 3,384 & 23,879 & $2.54 \times 10^6$ & N/A & N/A & N/A & N/A & -- \\
c499    & 10    & 236    & 397 & 55,585 & N/A & N/A & N/A & -- \\
c880    & 67    & 1,601  & 1,903 & $8.22 \times 10^6$ & Timeout & Timeout & Timeout & A=881.953, B=1.75536 \\
c1355   & 59    & 636    & $5.67 \times 10^6$ & $1.96 \times 10^9$ & Timeout & Timeout & Timeout & A=21.7956, B=3.66316 \\ 
c1908   & 13    & 294    & 12,594 & $1.33 \times 10^7$ & Timeout & Timeout  & Timeout & A=0.00174972, B=4.59814 \\
c2670   & 273   & 1,570  & 8,912 & $2.90 \times 10^5$ & Timeout & Timeout & Timeout & A=1036.51, B=1.82463 \\ 
c3540   & 1,215 & 5,991  & $8.69 \times 10^6$ & $4.98 \times 10^8$ & Timeout & Timeout & Timeout  & A=4.0758, B=3.72457 \\
c5315   & 162   & 4,869  & 6,650 & $1.22 \times 10^9$ & Timeout & Timeout  & Timeout & A=0.0217371, B=4.95008 \\
c7552   & 11    & 124    & 1,558 & $2.57 \times 10^5$ & $1.15 \times 10^9$ & Timeout & Timeout  & A=9191.65, B=1.10724 \\ \bottomrule
\end{tabular}
\end{table}


\begin{table}[t]
\footnotesize
\centering
\caption{Area overhead of SCs versus number of used MCs.}
\label{area_overhead}
\begin{tabular}{@{} *8c @{}}
\toprule 
\#MCs   & N=1 & N=2 & N=3 & N=5 & N=10 & N=15 & N=20 \\
\midrule
\#Gates & 6   & 11  & 16  & 26  & 51   & 76   & 101  \\ \bottomrule
\end{tabular}
\end{table}

The number of cycles created in ISCAS-85 benchmarks, when adding N=1, 2, 3, 5, 10, 15 and 20 MCs of size 7 (i.e., 7 gates in a cycle), while building a super cycle is reported in Table \ref{table_of_SC}. As illustrated, the number of cycles created from implementing a super cycle is primarily a function of the number of inserted feedbacks and secondarily a function of the topology of starting netlist. As expected, the number of cycles is aggressively increased by the addition of each MC to the SC. Since the number of created cycles is also a function of topology, no single equation could predict the exact exponential growth for all benchmarks. Hence, using curve fitting techniques, the number of created cycles in each netlist as a function of the number of feedbacks is also reported in this table. As expected in all cases we see an exponential relation between the number of feedbacks and the number of created cycles. In smaller circuits like c432 and c499, there were not enough gates to create the required MCs for larger super cycles, hence, N/A is reported. For timeout entries, the number of cycles was not determined after 10 hours of execution on our server node. For executions resulted in timeout we also confirmed that initiating the CycSAT with incomplete NC clauses traps the SAT solver in an infinite loop. As illustrated in Table. \ref{table_of_SC}, this is the case for most of the circuits with more than 10 MCs. The area overhead for building the SC in terms of the number of needed switches depends on the number of MCs and the number of gates in each MC. The area overhead for having various number of MCs of size 7 gates when building a SC is reported in Table \ref{area_overhead}. In short, as the number of inserted feedbacks increases, the pre-processing step of CycSAT faces an exponential increase in its runtime. Hence, by introducing a reasonable number of feedbacks using a methodology that exponentially increases the number of cycles (such as SC or LFN as proposed in this paper), the netlist could be protected against both SAT and CycSAT attacks by means of cyclic obfuscation. 


\begin{table}[h]
\footnotesize
\centering
\caption{Number of cycles in c1908 circuit after inserting different number of cycles and SR latches.}
\label{cycle_sr}
\setlength\tabcolsep{1pt} 
\begin{tabular}{@{} *7c @{}}
\toprule
\hspace{0.5pt} \backslashbox[45pt]{\#MCs}{\#SR-L} & 0 & 1 & 2 & 5 & 10 & 20 \\
\midrule
0    & 0          & 5 & 30 & 245 & 7,675 & $7.35 \times 10^6$ \\
1    & 13         & 189 & 9,875 & $4.37 \times 10^6$ & $1.78 \times 10^7$ & Timeout \\
2    & 294        & 7,574 & $8.71 \times 10^6$  & Timeout & Timeout & Timeout \\
3    & 12,594     & $6.82 \times 10^5$ & $1.42 \times 10^7$ & Timeout & Timeout & Timeout \\
5    & $1.33 \times 10^7$ & $2.28 \times 10^7$ & Timeout & Timeout & Timeout & Timeout \\ 
10   & Timeout    & Timeout & Timeout & Timeout & Timeout & Timeout \\ 
\bottomrule
\end{tabular}
\end{table}

Table \ref{cycle_sr} illustrates the strength of cyclic obfuscation when a netlist (c1908 in ISCAS-85) is first cyclified and then is subjected to cyclic obfuscation. For netlist cyclification, the input-dependency-based technique in section \ref{idbo} is used, and for exponentially increasing the number of cycles with respect to number of feedbacks, the SC approach was used. After adding 5 SR-latches and only two MCs, list of cycles could not be generated under 10 hour time limit. This will again prevent the initiation of SAT-attack because cycle avoidance clauses could not be generated. In short, cyclification of a netlist before subjecting it to cyclic obfuscation could increase the runtime of CycSAT pre-processing step exponentially. 

\section{Conclusion}\label{conclusion}
By introducing cycles in a netlist, the straightforward SAT-attack would be trapped in an infinite loop while CycSAT can solve such obfuscation problems. However, the problem with CycSAT is the runtime of its pre-processing step for generating the cycle avoidance clauses, which grow as a linear function of the number of cycles in the netlist. As a mean of defense, we introduced several techniques for exponentially increasing the number of cycles with respect to the number of inserted feedbacks. This, in turn, resulted in an exponential increase in runtime of CycSAT pre-processing step, disabling the SAT attack to be carried in a reasonable amount of time. Based on this study, the cyclic obfuscation when properly implemented poses an exponential runtime on CycSAT attack with respect to the number of inserted feedbacks. Hence, CycSAT or existing SAT-attacks are not an effective mean for breaking cyclic obfuscation.

\bibliographystyle{ACM-Reference-Format}
\vspace{-5pt}
\bibliography{refs}

\end{document}